\documentclass[10pt]{elsarticle}
\usepackage{mathrsfs}
\usepackage{graphicx}
\usepackage{amssymb}
\usepackage{amsmath}
\usepackage{amsfonts}
\usepackage{natbib}
\usepackage{lineno}
\usepackage{bm}
\usepackage{xr}
\usepackage{pdfpages}

\newcounter{dummy}
\newtheorem{corollary}[dummy]{Corollary}
\newtheorem{lemma}[dummy]{Lemma}
\newtheorem{theorem}[dummy]{Theorem}

\newcommand{\be}{\begin{equation}} \newcommand{\ee}{\end{equation}}
\newcommand{\bd}{\begin{displaymath}} \newcommand{\ed}{\end{displaymath}}
\newcommand{\ba}{\begin{align}} \newcommand{\ea}{\end{align}}
\newcommand{\baa}{\begin{align*}} \newcommand{\eaa}{\end{align*}}
\newcommand{\ben}{\begin{enumerate}} \newcommand{\een}{\end{enumerate}}
\newcommand{\bi}{\begin{itemize}} \newcommand{\ei}{\end{itemize}}

\newcommand{\E}[1]{\operatorname{E}\left[ #1 \right]}
\newcommand{\Var}[1]{\operatorname{Var}\left[ #1 \right]}

\journal{Journal of Theoretical Biology}

\begin{document}

\begin{frontmatter}

\title{Time to a single hybridization event in a group of species with unknown ancestral history}

\author[CTHGU]{Krzysztof Bartoszek}
\author[CTHGU,GU]{Graham Jones}
\author[GU]{Bengt Oxelman}
\author[CTHGU]{Serik Sagitov\corref{cor1}}
\ead{serik@chalmers.se}
\cortext[cor1]{Corresponding author}
\address[CTHGU]{Mathematical Sciences, Chalmers University of Technology and the University of Gothenburg, Gothenburg, Sweden}
\address[GU]{Department of Biological and Environmental Science, University of Gothenburg, Gothenburg, Sweden}

\begin{abstract}
We consider a stochastic process for the generation of species which combines a Yule process with a simple model for hybridization between pairs of co-existent species. We assume that the origin of the process, when there was one species, occurred at an unknown time in the past, and we condition the process on producing $n$ species via the Yule process and a single hybridization event. We prove results about the distribution of the time of the hybridization event. In particular we calculate a formula for all moments, and show that under various conditions, the distribution tends to an exponential with rate twice that of the birth rate for the Yule process.
\end{abstract}

\begin{keyword}
Theoretical phylogenetics \sep Yule tree \sep Polyploidy  \sep Uncertainty in phylogeny
\MSC[2010] 60J70 \sep 60J85 \sep 62P10 \sep 92B99
\end{keyword}

\end{frontmatter}


\section{Introduction}

Hybridization has an important role in the evolution of new species \cite{Arnold,Mallet}. In phylogenetic analysis, there is an increasing interest in dealing with this issue \cite{Kubatko,JSO,YDN}. The usual phylogenetic tree is replaced by a phylogenetic network \cite{HRS}, and in a Bayesian approach, a prior for the network is needed \cite{JSO}. Very little is known about suitable prior distributions for the topology and node times for such networks. This paper represents an attempt to understand the situation better, and provides some justification for using an exponential distribution as a prior for the hybridization time.  

The particular biological motivation for this study originates from a theoretical question on the evolution of polyploidy in plants. Polyploids can arise from within a single species (autoployploids) or via hybridization between two species (allopolyploids) in which the genomes of the two parental species are both present in the hybrid. For example, suppose it is known that a tetraploid species of interest resulted from a hybridization between a pair of diploid species which are ancestral to a clade of $n$ extant species. The following question arises: what can we say about the time of the hybridization event prior to a phylogenetic analysis of the genetic data? 

The same question can be applied to homoploid hybridization, in which there is a hybridization but no change in ploidy. However we will refer to the allopolyploid case above, since the species produced by the Yule process and the hybrids can be conveniently called diploids and tetraploids.
\begin{figure}
\centering
\includegraphics[width=0.8\textwidth]{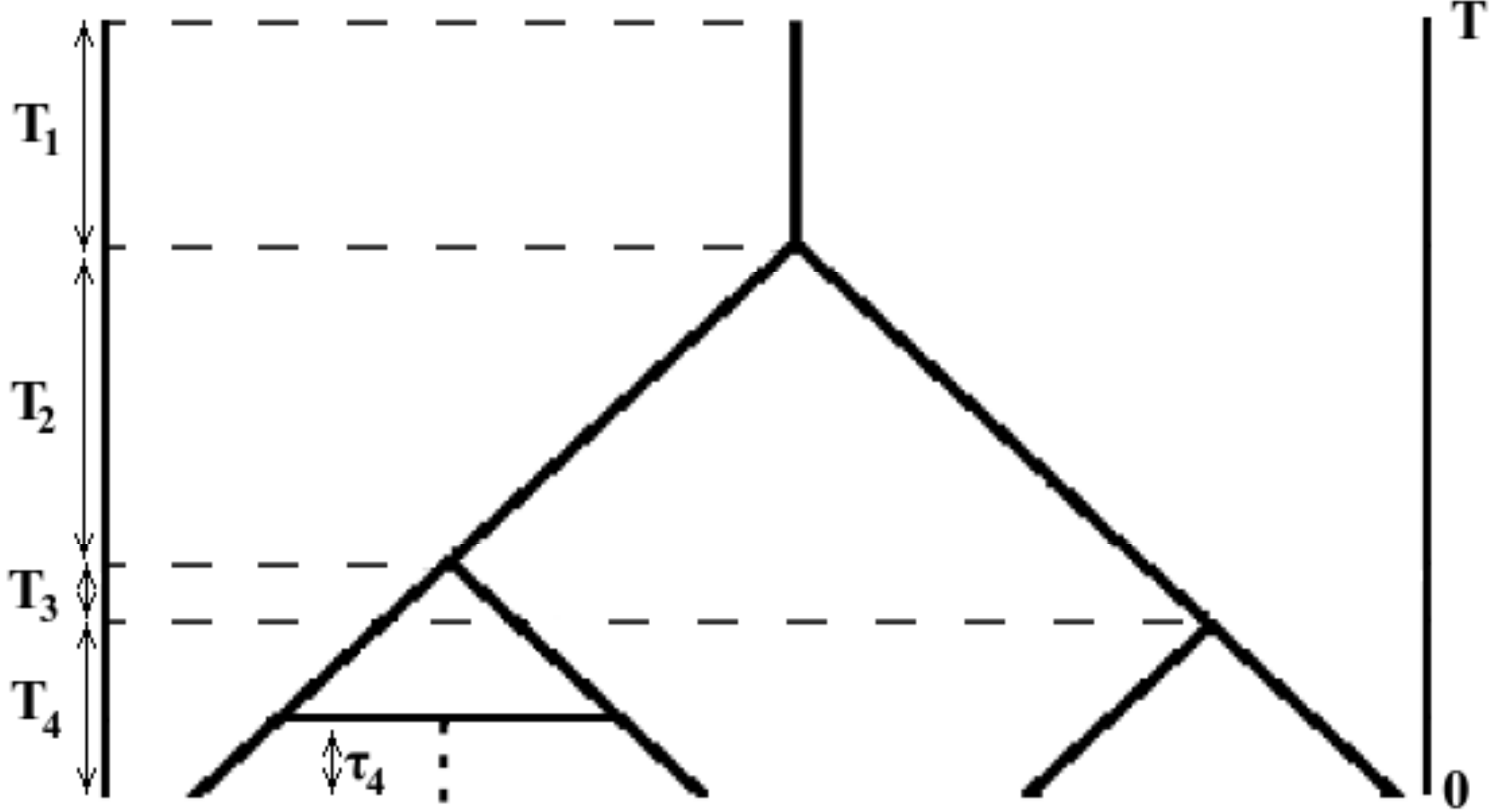}
\caption{Main time characteristics of the  of the conditional Yule tree for $n=4$ species with one hybridization: $T$ is the time to origin, $T_1,\dots,T_4$ are inter-speciation times, and $\tau_4$ is the time to hybridization.}
\label{fig0}
\end{figure}

We assume a Yule model with the speciation rate $\lambda$ conditioned on $n$ extant species and model the hybridization events by a Poisson process with intensity $\beta$ giving the number of hybridizations per pair of
coexisting diploid species per unit of time calibrated by $\lambda$. This means that if there are $k$ coexisting diploid species during a time period $t$, then the number of hybridizations $N_k(t)$ during this period has a Poisson distribution
\begin{equation}\label{Poi}
 P(N_k(t)=j)={\beta{k\choose2}t\over j!}e^{-\beta{k\choose2}t}, \ j=0,1,2,\ldots
\end{equation}
with expectation
\begin{equation}\label{Poim}
 E(N_k(t))=\beta{k\choose2}t.
\end{equation}

Counting time backwards, let $T_k$ stand for the time between two consecutive speciation events during which the Yule tree had $k$ branches, $k=2,\ldots,n$, see Fig. \ref{fig0}.
In the conditioned Yule model setting (a random phylogeny for $n$ extant species under the assumption of an improper uniform prior for the time of origin \cite{GT}) the times $(T_2,\ldots,T_n)$ are independent and exponentially distributed random variables with parameters $(2\lambda,\ldots,n\lambda)$ respectively. Replacing $t$ by $T_k$ in formula \eqref{Poim} and writing $N_k=N_k(T_k)$ gives
$$E(N_k)=\beta{k\choose2}E(T_k)=\gamma(k-1),$$
where the compound parameter $\gamma=\frac{\beta}{2\lambda}$ can be understood as a relative hybridization rate. 
Thus averaging over possible species trees results in the mean total number of hybridizations $N=N_2+\ldots+N_n$ being
\begin{equation}\label{Poit}
 E(N)=\gamma {n\choose 2}.
\end{equation}

The main finding of this paper is that the distribution of the time $\tau_n$ to a single hybridization event can be approximated by an exponential distribution with parameter $2\lambda$. This obtained by showing, see Corollary \ref{cor}, that $r$-th moment of $2\lambda\tau_n$ converges to $r!$ which is the $r$-th moment of an exponential distribution with parameter 1. Our simulations show that  even for moderate values of $n$ and reasonable values of $\gamma$ the exponential approximation for the time to hybridization seem to be satisfactory.

\section{The single hybridization condition}\label{Ssh}
Given that there was exactly one hybridization event, $N=1$, we denote by $\tau_n$ the time to hybridization counted backwards from the time of observation. If $N=0$ or $N\ge2$, we put $\tau_n=\infty$. In this section we show among other things that the single hybridization condition has probability
\begin{equation}\label{prob}
P(\tau_n<\infty)=G_n\prod_{i=1}^{n-1}\frac{1}{1 + i\gamma},
\end{equation}
where $G_n=\sum_{k=1}^{n-1} \frac{k\gamma}{1 +k\gamma}$. 
Observe that if  $\tau_n<\infty$, then for some  $\kappa_n\in\{2,\ldots,n\}$ hybridization occured during the period when there were $\kappa_n$ ancestral species.
\begin{lemma}\label{le1}
 For any $2\le k\le n<\infty$
 \begin{align*}
P(\kappa_n=k|\tau_n<\infty)=G_n^{-1} \frac{(k-1)\gamma}{1 +(k-1)\gamma}.
\end{align*}
\end{lemma}
{\sc Proof of Lemma \ref{le1}.} Replacing $t$ by $T_k$ in the right hand side of \eqref{Poi} yields
\begin{align*}
P(N_k=0|T_2,\ldots,T_n)&=e^{-\beta{k\choose2}T_k},\\
P(N_k=1|T_2,\ldots,T_n)&=\beta{k\choose2}T_ke^{-\beta{k\choose2}T_k},
\end{align*}
and since 
$$\{\kappa_n=k\}=\{N_n=0,\ldots,N_{k+1}=0,N_k=1,N_{k-1}=0,\ldots,N_{2}=0\},$$
we obtain
\begin{align}\label{kap}
P(\kappa_n=k|T_2,\ldots,T_n)&=\beta{k\choose2}T_k\prod_{i=2}^ne^{-\beta{i\choose2}T_i},
\end{align}
and therefore
\begin{align}\label{kax}
P(\kappa_n=k)&={\beta{k\choose2}\over \lambda k+\beta{k\choose2}}\prod_{i=2}^n{\lambda i\over \lambda i+\beta{i\choose2}}=\frac{(k-1)\gamma}{1 +(k-1)\gamma}\prod_{i=1}^{n-1}\frac{1}{1 + i\gamma}.
\end{align}
Summing over $k=2,\ldots,n$ we arrive at \eqref{prob}, then the assertion of Lemma \ref{le1} follows by dividing the last expression by \eqref{prob}. \\

If we assume that $n$ diploids and a single hybridization have been observed, then we can apply two basic methods of estimation for the plausible value of the key parameter $\gamma$. The method of moments estimate $\tilde\gamma_n=1/{n\choose2}$ is immediately obtained from  \eqref{Poit} by substituting the observed value $N=1$ for $E(N)$. 
We can also treat the expression for $P(\tau_n<\infty)$ in \eqref{prob} as a likelihood function for $\gamma$
\[L(\gamma)=\prod_{i=1}^{n-1}\frac{1}{1 + i\gamma}\sum_{k=1}^{n-1} \frac{k\gamma}{1 +k\gamma}\] 
and from it find a maximum likelihood estimate $\hat\gamma_n$. It turns out that for large $n$ the two estimates are close in value
\begin{align}\label{mle}
\hat\gamma_n\ge\frac{2}{n(n-1)} \mbox{  for } n\ge2, \mbox{ and } \hat\gamma_n\le \frac{2}{n(n-3)}  \mbox{  for } n \geq 4.
\end{align}

To show \eqref{mle} we observe first that the equation $L'(\hat\gamma)=0$ for $\hat\gamma_n$ takes the form
\begin{align}\label{ml}
A(\hat\gamma) = \hat\gamma B(\hat\gamma)^2,
\end{align}
where
$$A(x) = \sum_{k=1}^{n-1} \frac{k}{(1 +kx)^2} \mbox{\ \ and\ \ } B(x) = \sum_{k=1}^{n-1} \frac{k}{1 +kx}.$$
By the Cauchy-Schwarz inequality we have
\begin{align*}
 B(x)^2  &= \Big(\sum_{k=1}^{n-1}\sum_{i=1}^{k} \frac{1}{1 +kx}1_{\{i\ge1\}}\Big)^2\\&\le\sum_{k=1}^{n-1}\sum_{i=1}^{k} \Big(\frac{1}{1 +kx}1_{\{i\ge1\}}\Big)^2\times \sum_{k=1}^{n-1}\sum_{i=1}^{k} \Big(1_{\{i\ge1\}}\Big)^2
=A(x){n(n-1)\over2},
\end{align*}
which together with \eqref{ml} yields $1\le \hat\gamma_n \frac{n(n-1)}{2}$. On the other hand, since for $x\ge0$ 
$$B(x)\ge A(x)\quad \text{and}\quad (1 +nx)B(x) \geq \frac{{n(n-1)}}{2},$$
it follows from \eqref{ml} that $1+n\hat\gamma_n \ge\hat\gamma_n \frac{n(n-1)}{2}$ and $1\ge\hat\gamma_n \frac{n(n-3)}{2}$.

\section{Exact formula for any moment of $\tau_n$}

\begin{lemma}\label{rth}
 For any $r\ge1$
\begin{align*}
E\left( \tau_n^r | \tau_n<\infty \right) &= 
G_n^{-1} \frac{r!}{\lambda^r} \sum_{k=1}^{n-1} \frac{k\gamma}{1 + k\gamma} \sum_{i_1=k}^{n-1}\sum_{i_2=i_1}^{n-1}\dots\sum_{i_r=i_{r-1}}^{n-1}d_{i_1}\cdots d_{i_r},
\end{align*}
where $d_j=(1+j)^{-1}(1+\gamma j)^{-1}$.
\end{lemma}
{\sc Proof of Lemma \ref{rth}.}
Under the Poisson model for the flow of hybridization events 
 \begin{align}
\tau_n&=X+\sum_{j=\kappa_n+1}^n T_j,
\label{repr}
\end{align}
where $X$ is a random variable uniformly distributed on $[0, T_{\kappa_n}]$. Thus
\begin{align*}
E\left( \tau_n^r | \kappa_n=k \right) &= E\left( \Big(X + \sum_{j=k+1}^{n} T_{j}\Big)^r \bigg| \kappa_n=k \right)\\
 &=E\left( \sum_{\alpha} \frac{r!}{\alpha_k! \cdots \alpha_{n}!} 
     X^{\alpha_k}\prod_{i=k+1}^n T_{i}^{\alpha_{i}} \bigg| \kappa_n=k \right),
\end{align*}
where the sum is over all vectors $\alpha = (\alpha_k, \dots \alpha_{n})$ of non-negative integers with sum $r$. Next we take such an 
$\alpha$ and calculate the expectation of
\begin{equation*}
M_{k,\alpha}=  X^{\alpha_k} \prod_{i=k+1}^n T_{i}^{\alpha_{i}} \cdot 1_{\{\kappa_n=k \}} .
\end{equation*}
We have in view of \eqref{kap}
\begin{align*}
E(M_{k,\alpha}) &= E \left( \left(T_k^{-1}\int_0^{T_k}x^{\alpha_k}dx\right)\times \prod_{i=k+1}^n T_{i}^{\alpha_{i}}  \times \beta{\binom{k}{2}} T_k 
\prod_{j=2}^n e^{-\beta{\binom{j}{2}}T_j} \right) \\
&= \prod_{j=2}^{k-1} E \Big( e^{-\beta{\binom{j}{2}}T_j} \Big) 
\beta{\binom{k}{2}}E \Big( { T_k^{1+\alpha_k}e^{-\beta{\binom{k}{2}}T_k}\over 1+\alpha_k}\Big)
\prod_{i=k+1}^n E \Big(T_i^{\alpha_{i}}e^{-\beta{\binom{i}{2}}T_i}\Big) \\
&= \prod_{j=2}^{k-1} \frac{1}{1 + (j-1)\gamma} \times
\gamma (k-1) \frac{ \alpha_k! d_{k-1}^{\alpha_k} \lambda^{-\alpha_k} }{ (1 + (k-1)\gamma)^2  } \times\prod_{i=k+1}^n \frac{\alpha_{i}! d_{i-1}^{\alpha_{i}} \lambda^{-\alpha_{i}}}{1 + (i-1)\gamma}.
\end{align*}
Recalling \eqref{kax} we deduce
\begin{align*}
E(M_{k,\alpha})  &=\frac{(k-1)\gamma}{1 +(k-1)\gamma}\Big( \prod_{i=1}^{n-1}\frac{1}{1 + i\gamma}\Big)
\Big(\lambda^{-r} \prod_{i=k}^n \alpha_{i}! d_{i-1}^{\alpha_{i}}\Big)\\
 &= P(\kappa_n=k) \lambda^{-r} \prod_{i=k}^n \alpha_{i}! d_{j-1}^{\alpha_{i}},
\end{align*}
which implies
\begin{align*}
E\left( \tau_n^r | \kappa_n=k \right) &= 
\frac{r!}{\lambda^r} \sum_{\alpha} \prod_{i=k}^n d_{i-1}^{\alpha_{i}}
= \frac{r!}{\lambda^r} \!\!\!\!\!\!\!\!\!\! \sum_{\substack{i_1, \dots, i_r \\ k-1 \leq i_1 \dots \leq i_r \leq n-1}} \prod_{j=1}^{r} d_{i_j}.
\end{align*}
Now to finish the proof of Lemma \ref{rth} it remains to apply Lemma \ref{le1}.\\

In particular, for $r=1$ and $r=2$ Lemma \ref{rth} gives
\begin{align}
m_n:=\E{\tau_n|\tau_{n}<\infty}&=\lambda^{-1} G_n^{-1}\sum_{k=1}^{n-1} \frac{k\gamma}{1 +k\gamma}\sum_{j=k}^{n-1}d_j,
\label{exp}
\end{align}
and
\begin{align*}
\E{\tau_n^2|\tau_{n}<\infty}&=2\lambda^{-2}G_n^{-1}\sum_{k=1}^{n-1} \frac{k\gamma}{1 +k\gamma}\sum_{j=k}^{n-1}  \sum_{l=j}^{n-1}d_jd_l \nonumber\\
&=\lambda^{-2}G_n^{-1}\sum_{k=1}^{n-1} \frac{k\gamma}{1 +k\gamma}\Big\{\Big(\sum_{j=k}^{n-1} d_j\Big)^2+\sum_{j=k}^{n-1} d_j^2\Big\},
\end{align*}
implying
\begin{align}
\Var{\tau_n|\tau_{n}<\infty}&=G_n^{-1}\sum_{k=1}^{n-1} \frac{k\gamma}{1 +k\gamma}\Big\{\Big(\lambda^{-1}\sum_{j=k}^{n-1} d_j-m_n\Big)^2+\lambda^{-2}\sum_{j=k}^{n-1} d_j^2\Big\}.
\label{var}
\end{align}
Here we have used the following observation: in terms of $Y_n:=\lambda^{-1}\sum_{j=\kappa_n-1}^{n-1} d_j$ we have $m_n=\E{Y_n}$ and
\begin{align*}
\Var{\tau_n|\tau_{n}<\infty}&=\E{Y_n^2}-m_n^2+\lambda^{-2}G_n^{-1}\sum_{k=1}^{n-1} \frac{k\gamma}{1 +k\gamma}\sum_{j=k}^{n-1} d_j^2\\
&=\E{(Y_n-m_n)^2}+\lambda^{-2}G_n^{-1}\sum_{k=1}^{n-1} \frac{k\gamma}{1 +k\gamma}\sum_{j=k}^{n-1} d_j^2.\end{align*}

\section{Convergence to an exponential distribution}
For $2\le k\le n<\infty$ and any natural number $r$ define $\eta_{\gamma,k,n}$ and $\zeta_{\gamma,n,r}$ by
 \begin{align*}
P(\kappa_n\le k|\tau_n<\infty)&= {k(k-1)\over n(n-1)}(1+\eta_{\gamma,k,n}),\\
\E{(2\lambda\tau_{n})^r\vert \tau_{n}<\infty} &=r!(1-\zeta_{\gamma,n,r}).
\end{align*}

\begin{figure}
\centering
\includegraphics[width=0.35\textwidth]{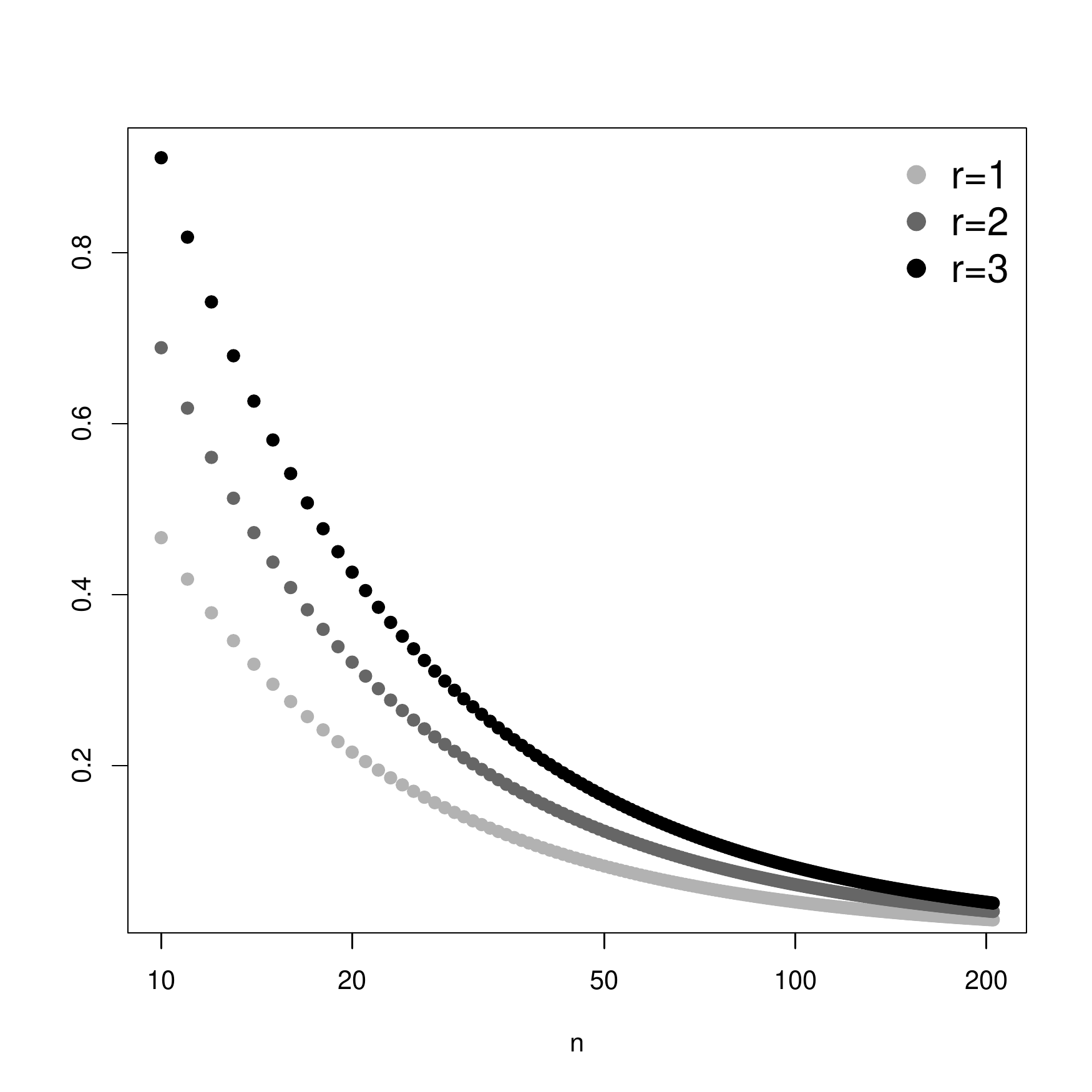}
\caption{The upper bound in  \eqref{wok}  for $\gamma={2\over n(n-1)}$ equals
${2\over n(n-1)}+{2(r+1)\over n-1}$. This upper bound is illustrated by plotting three functions of $n$ for the first three moments $r=1,2,3$.
}
\label{fig10}
\end{figure}

\begin{theorem}\label{the}
For $2\le k\le n<\infty$ and $r\ge1$ the following bounds are valid
 \begin{align}
-k\gamma&\le\eta_{\gamma,k,n}\le n\gamma,\label{wk}\\
\label{wok}
0&\le\zeta_{\gamma,n,r}\le(1+(r+1)n)\gamma.
\end{align}
 \end{theorem}
The discussion in the end of Section \ref{Ssh} concerning \eqref{mle} showed that it is important to consider the values of $\gamma$ close to ${2\over n(n-1)}$. In Figure \ref{fig10} we plotted the upper bounds in  \eqref{wok}  with $\gamma={2\over n(n-1)}$
as functions of $n$ for the first three moments $r=1,2,3$.

\begin{corollary}\label{cor}
Uniformly over all $(k,n)$ such that $2\le k\le n<\infty$
 \begin{align}\label{wk1}
P(\kappa_n= k|\tau_n<\infty)\to {2(k-1)\over n(n-1)},\quad n\gamma\to0.
\end{align}
Moreover, as $n\gamma  \rightarrow 0$ for any fixed natural number $r$
\begin{equation}\label{wok1}
\E{(2\lambda\tau_{n})^r\vert \tau_{n}<\infty}  \to r!
\end{equation}
uniformly over $\lambda\in(0,\infty)$. 
 \end{corollary}
 
 Corollary \ref{cor} is a straightforward consequence of  Theorem \ref{the} proved next. Note that 
in the case $\gamma={2\over n(n-1)}$ the condition $n\gamma  \rightarrow 0$  is equivalent to $n\to\infty$.  

{\sc Proof of Theorem \ref{the}}.
According to Lemma \ref{le1}
 \begin{align*}
P(\kappa_n\le k|\tau_n<\infty)=G_k/G_n,
\end{align*}
and \eqref{wk} follows from
\begin{equation}\label{gin}
{1\over 1+n\gamma}\gamma \binom{n}{2}\le G_{n}\le{1\over 1+\gamma}\gamma \binom{n}{2}.
\end{equation}

To prove \eqref{wok} observe first that
\begin{align*}
\sum\limits_{k=1}^{n-1}k\sum\limits_{i_{1}=k}^{n-1}\ldots&\sum\limits_{i_{r}=i_{r-1}}^{n-1}\left(\frac{1}{1+i_{1}}\cdots\frac{1}{1+i_{r}}\right)\\
&=\sum\limits_{i_{r}=1}^{n-1}
\sum\limits_{i_{r-1}=1}^{i_{r}}\ldots\sum\limits_{i_{1}=1}^{i_{2}}\left(\frac{1}{1+i_{1}}\cdots\frac{1}{1+i_{r}}\right)\sum\limits_{k=1}^{i_{1}}k\\ 
&=2^{-1}\sum\limits_{i_{r}=1}^{n-1}
\ldots\sum\limits_{i_{2}=1}^{i_{3}}\left(\frac{1}{1+i_{2}}\cdots\frac{1}{1+i_{r}}\right)\sum\limits_{i_{1}=1}^{i_{2}}i_{1}\\ 
& =
2^{-r}\binom{n}{2}.
\end{align*}
Clearly, for any $1\le k\le i_1\le i_2\le\dots\le i_r\le n-1$ we have
\begin{align*}
 {\gamma\over(1+n\gamma )^{r+1}}&{k\over (1+i_1)\cdots(1+i_r)} \\
 &\le {k\gamma\over 1+k\gamma}d_{i_1}\cdots d_{i_r}\le{\gamma\over(1+k\gamma)^{r+1}}{k\over (1+i_1)\cdots(1+i_r)}.
\end{align*}
Thus Lemma \ref{rth} yields
\begin{align*}
r!{\gamma \binom{n}{2}\over G_{n}(1+n\gamma )^{r+1}}\le\E{(2\lambda\tau_{n})^r\vert \tau_{n}<\infty} \le r!{\gamma \binom{n}{2}\over G_{n}},
\end{align*}
and  applying \eqref{gin} we get inequalities
\begin{align*}
{r!\over (1+\gamma)(1+n\gamma )^{r+1}}\le\E{(2\lambda\tau_{n})^r\vert \tau_{n}<\infty} \le r!
\end{align*}
resulting in \eqref{wok}.

\section{Simulation results and discussion}

\begin{figure}
\centering
\includegraphics[width=0.35\textwidth]{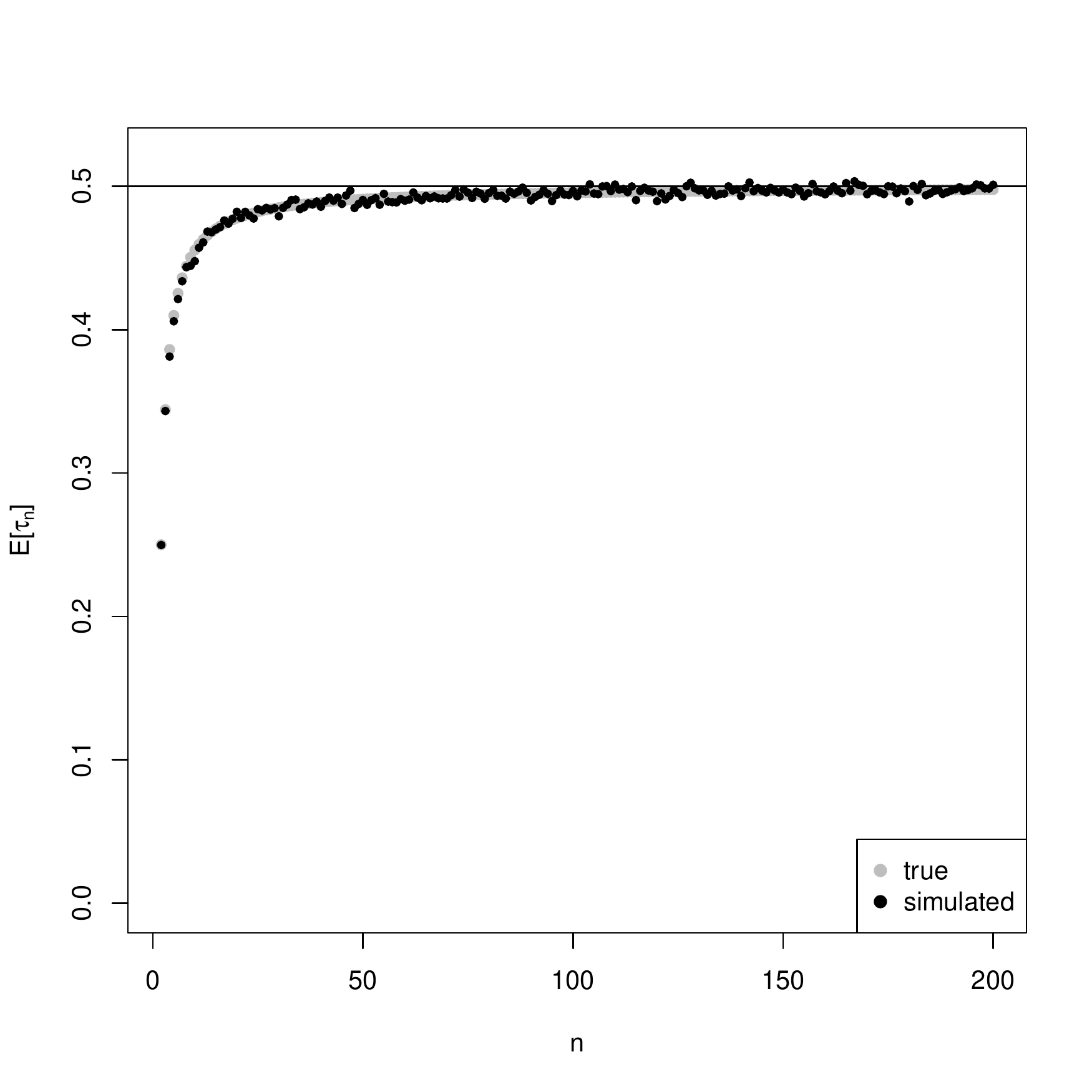}
\includegraphics[width=0.35\textwidth]{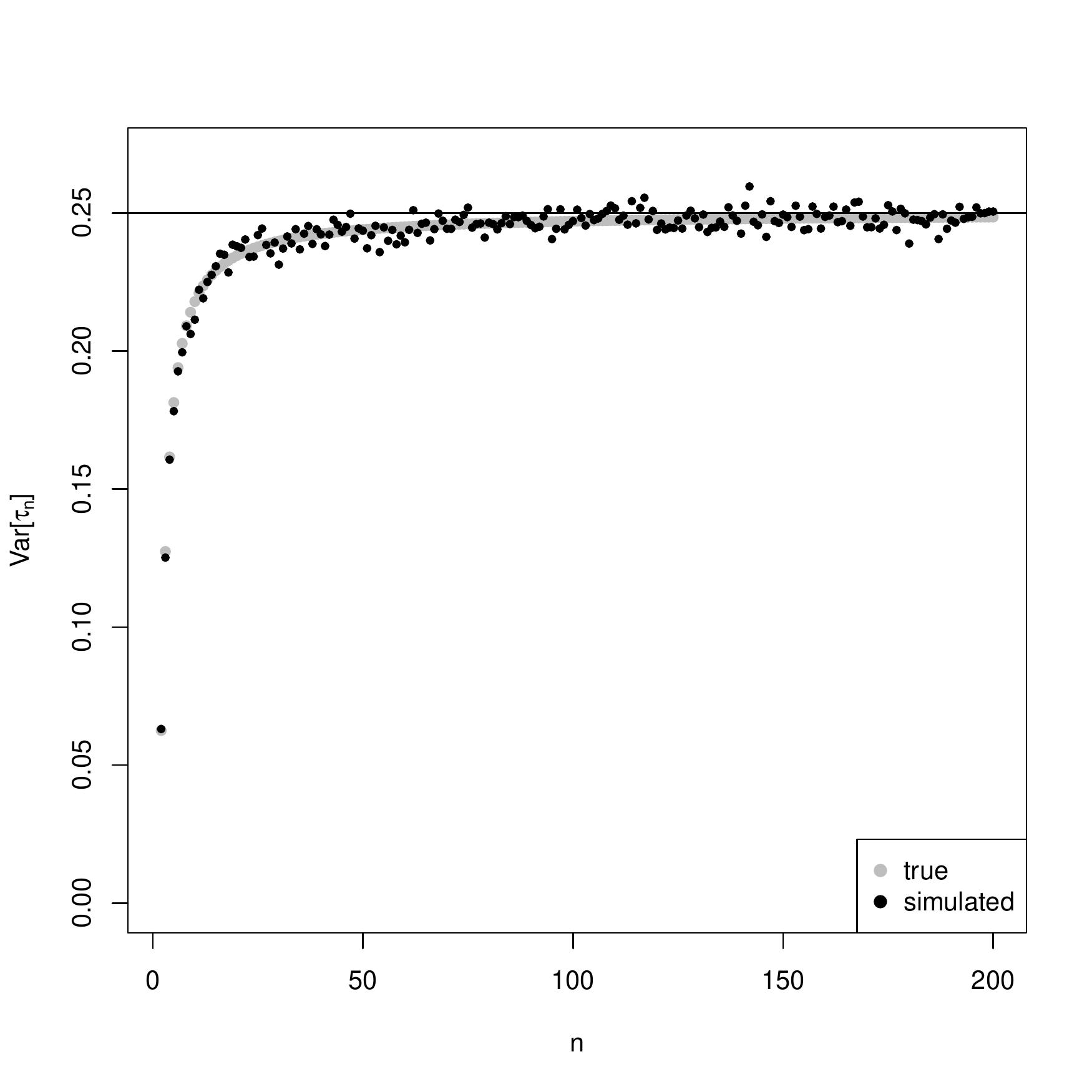}
\caption{Conditional mean and variance of $\tau_n$ as functions  \eqref{exp} and \eqref{var} of the number $n$ of candidate species. Simulations with $\lambda=1$ and $\beta={4\over n(n-1)}$ are compared to the analytical predictions.}
\label{fig1}
\end{figure}

We have checked and illustrated our analytical results using simulations. Our simulation algorithm is based on the following steps to obtain a single hybridization time:

\begin{description}

\item[Step 1] For ($k=1$ to $n$): $T_k \leftarrow $ sample from the exponential distribution with rate $k \lambda$.

\item[Step 2] For ($k=1$ to $n$): $r_k \leftarrow T_k \beta k(k-1) / 2$.

\item[Step 3] $R \leftarrow \sum_{k=2}^n r_k$.

\item[Step 4] $h \leftarrow $ sample from the Poisson distribution with mean $R$.

\item[Step 5] If ($h == 1$) then sample $k \in \{2,3,\dots,n\}$ with probability proportional to $r_k$, and then the hybridization time uniformly in the $k$th interval.

\end{description}

In Figure \ref{fig1} the mean and variance of $\tau_n$ as functions \eqref{exp} and \eqref{var} of the number $n$ of candidate species are drawn against the values obtained from simulations. Here $\lambda=1$ and $\gamma=\frac{\beta}{2}={2\over n(n-1)}$ with $n$ ranging from 2 to 200.

\begin{figure}
\centering
\includegraphics[width=0.32\textwidth]{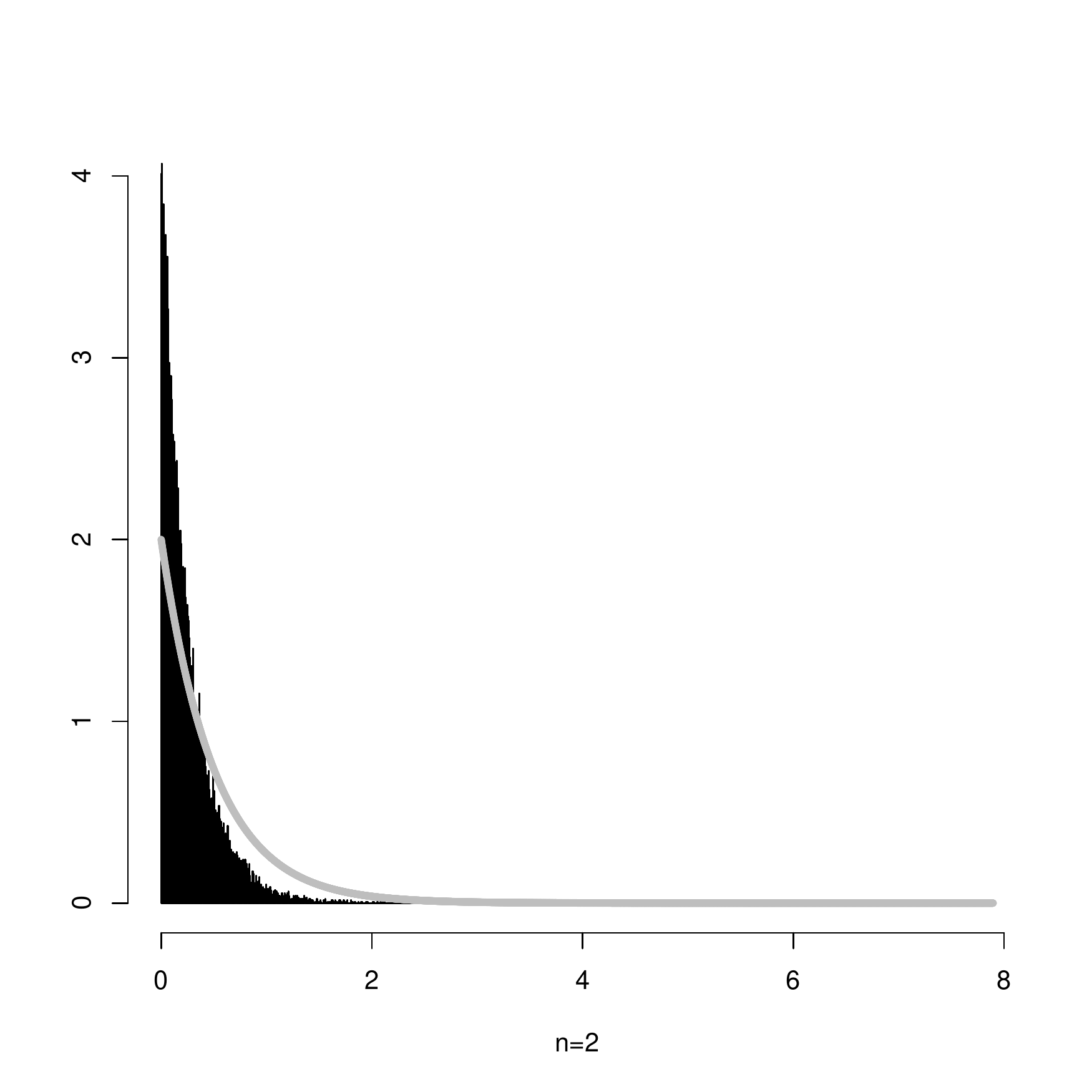}
\includegraphics[width=0.32\textwidth]{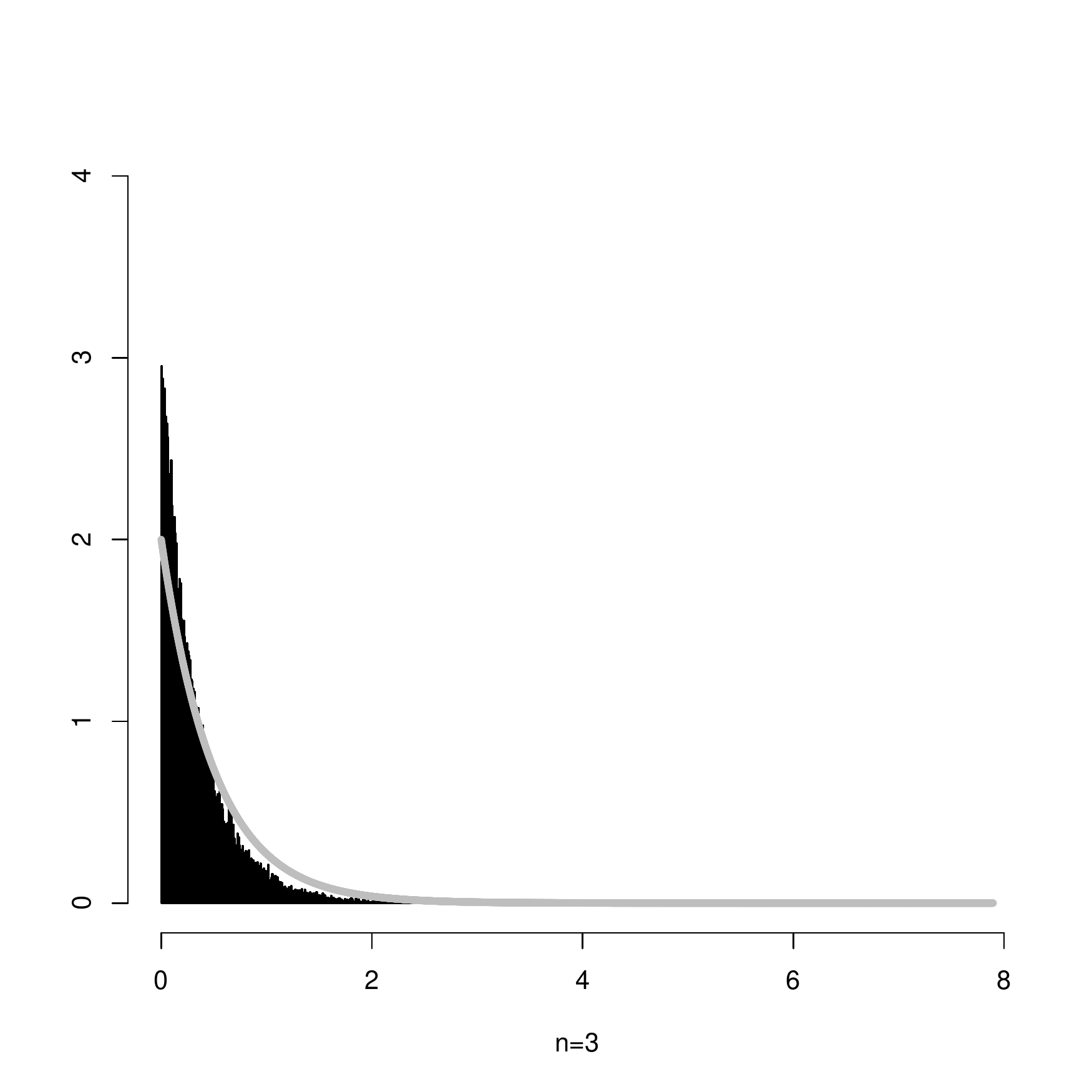}
\includegraphics[width=0.32\textwidth]{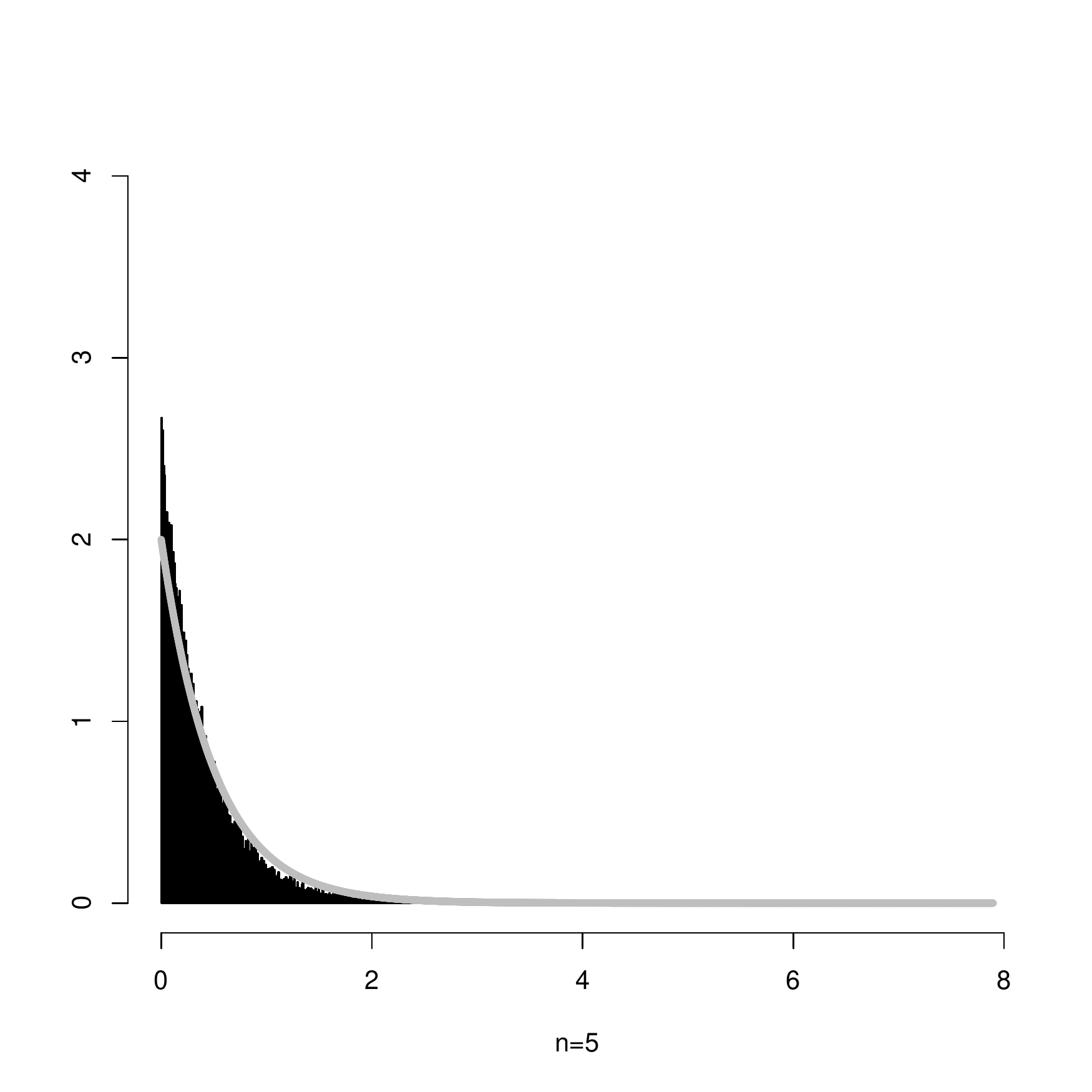}
\includegraphics[width=0.32\textwidth]{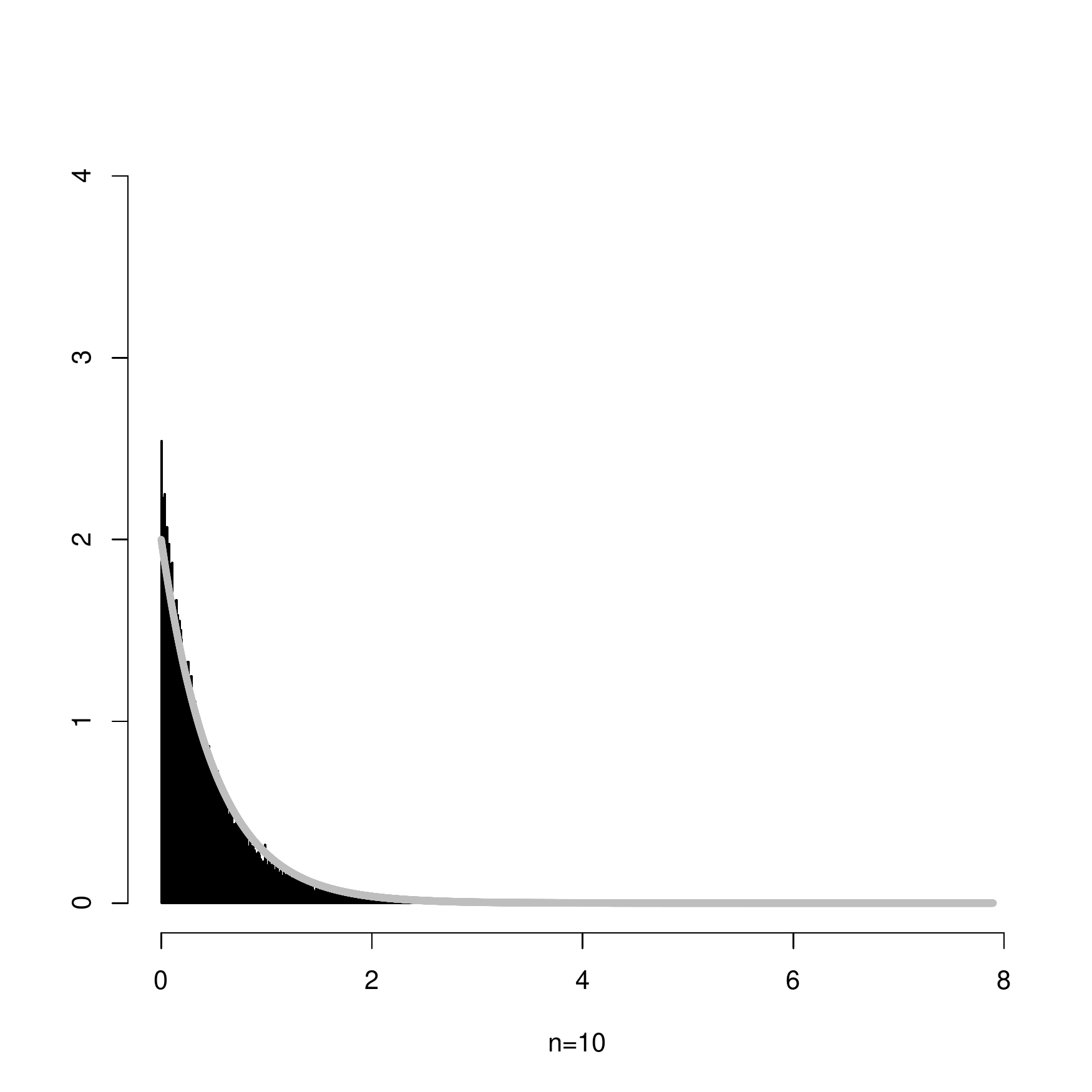}
\includegraphics[width=0.32\textwidth]{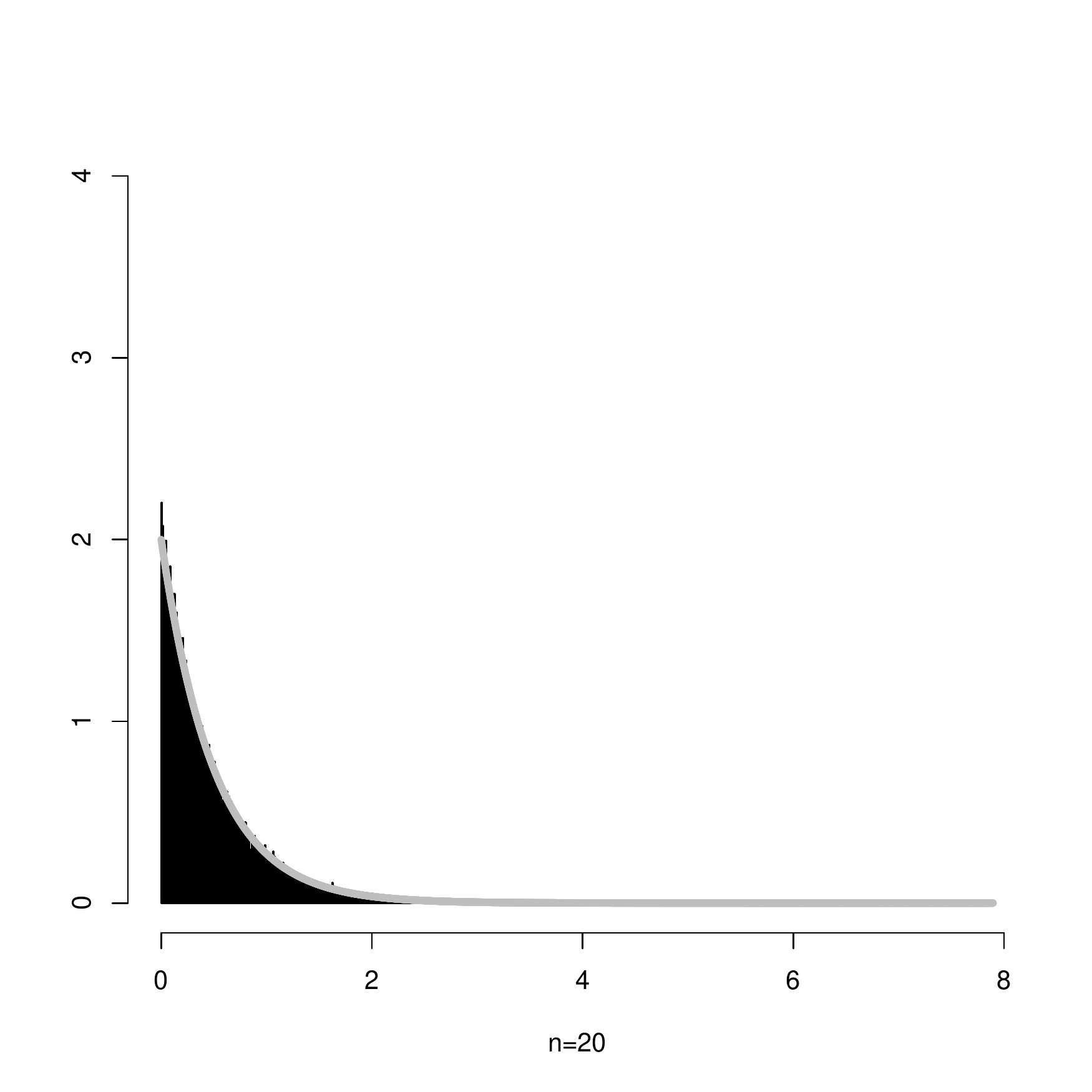}
\caption{Histograms for $\tau_n$ conditional on a single hybridization event for the number $n$ of candidate species. Left to right: $n=2,3,5,10,20$. Simulations with $\lambda=1$ and $\beta={4\over n(n-1)}$.}
\label{fig2}
\end{figure}

Figure \ref{fig2} shows simulated conditional distributions of $\tau_n$. We can see how the observed distribution profile approaches the exponential curve as $n$ increases from 2 to 20. 

The Yule model for the unknown species tree is not very realistic but it is a very convenient tool for phylogenetic calculations, see for example  \cite{BS}. Therefore, the presented here results should be viewed as just a starting point for the issues raised in this paper. More biologically relevant extensions of the model studied here should take into account the possibility  of hybridization between a pair of ancestral species of which either one or both have no direct descendants at present. To include extinct species in the analysis one can use the so-called conditioned {\it birth-death processes} developed in \cite{AP,GT} and successfully used as species tree models for various purposes, see for example \cite{SB}. An important additional parameter arising in this more general setting is the extinction rate $\mu$ for the ancestral species. 

An crucial biological feature missing in the classical birth-death processes modeling species trees is {\it geographical structure}. Obviously, the probability of hybridization is conditional on geographical proximity. This could presumably be taken into account by combining our model with some statistical biogeography model \cite{LRDS,R}.
Another desirable feature missing in the current analysis is the {\it decaying  hybridization rate}: the more divergent two species become, the less probable hybridization between them will be. Of course, one should not limit oneself only to one hybridization event allowing for {\it multiple hybridizations}. Furthermore, hybrids can speciate via ordinary speciation, and hybridizations between hybrids also occur, so these processes should be included in a general model.

\section*{Acknowledgments}
KB and GJ were supported by Centre for Theoretical Biology at the University of Gothenburg.
BO and SS were supported by Swedish Research Council grants 2009-5202 and 621-2010-5623. 
KB was supported by Stiftelsen f\"or Vetenskaplig Forskning och Utbildning i Matematik, 
Knut and Alice Wallenbergs travel fund, Paul and Marie Berghaus fund, Royal Swedish Academy of Sciences,
Wilhelm and Martina Lundgrens research fund.

\end{document}